\documentstyle[aps,version2,preprint,epsfig,rotating]{revtex}
\begin{document}
\draft
\preprint{OCIP/C 96-1}
\preprint{hep-ph/9608368}
\preprint{August 1996}
\begin{title}
Leptoquark Mass Limits \\
from Single Leptoquark Production at LEP and LEP200
\end{title}
\author{Michael A. Doncheski\thanks{Present address: 
Department of Physics, Pennsylvania State University, Mont Alto, PA 
17237 USA} and Stephen Godfrey}
\begin{instit}
Ottawa-Carleton Institute for Physics \\
Department of Physics, Carleton University, Ottawa CANADA, K1S 5B6
\end{instit}

\begin{abstract}
We investigate the  discovery potential for first generation
leptoquarks at the LEP and 
LEP200 $e^+e^-$ colliders.  We consider single leptoquark production via 
resolved photon contributions which offers a much higher kinematic 
limit than the more commonly considered leptoquark pair production 
process.  This process also has the advantage that it is independent 
of the leptoquark chirality, is almost insensitive to whether the 
leptoquark is scalar or vector, and only depends on the leptoquark 
charge.  We estimate that 
from the nonobservation of energetic $e-jet$ events,
the limits at LEP for all types of leptoquarks
are $M_{LQ} > 90$~GeV.
For the just completed 
$\sqrt{s}=161$~GeV run at LEP200 we estimate potential scalar 
leptoquark mass limits of
$M_{LQ} > 130$~GeV for $Q_{LQ}=-2/3, \; -4/3$
and $M_{LQ} > 148$~GeV for $Q_{LQ}=-1/3, \; -5/3$ and for vector LQ's
the limits are 140 and 154~GeV respectively.  
Ultimately, LEP200 should be 
able to achieve discovery limits for leptoquarks of $\sim 188$~GeV
for the $\sqrt{s}=190$~GeV 
runs.  In all cases we assume electromagnetic strength coupling.

\end{abstract}

There is considerable interest in the study of leptoquarks (LQs) --- colour
(anti-)triplet, spin 0 or 1 particles which carry both baryon and lepton 
quantum numbers.  Such objects appear in a large number of extensions 
of the standard model such as grand unified theories, technicolour, 
and composite models.  
Quite generally, the signature for leptoquarks is very
striking: a high $p_{_T}$ lepton balanced by a jet (or missing $p_{_T}$ 
balanced by a jet, for the $\nu q$ decay mode, if applicable).  
Searches for leptoquarks have been performed by the H1 \cite{H1} and 
ZEUS \cite{ZEUS}
collaborations at the HERA $ep$ collider, by the D0 \cite{D0} 
and CDF \cite{CDF} 
collaborations at the Tevatron $p\bar{p}$ collider, and by the ALEPH 
\cite{ALEPH}, 
DELPHI \cite{DELPHI}, L3 \cite{L3}, and OPAL\cite{OPAL} 
collaborations at the LEP $e^+e^-$ collider.  
The LQ limits obtained by the LEP experiments have for the most part 
considered LQ pair production although DELPHI considered single LQ 
production via the process $Z\to {\rm LQ} + e +jet$.  
In this letter we consider single 
leptoquark production in $e^+e^-$ collisions which utilizes the quark 
content of a Weizacker-Williams photon radiating off of one of the 
initial leptons\cite{DG1,DG2,eboli,leptoLN,HP,misc}.  
This process offers the advantage of a much higher 
kinematic limit than the LQ pair production process, is independent of 
the chirality of the LQ, and gives similar results for both scalar and 
vector leptoquarks.  The cross section depends on the LQ charge since 
the photon has a larger $u$ quark content than $d$ quark content and 
hence has a larger cross section for LQ's which couple to the $u$ 
quark than for LQ's which couple to the $d$ quark.

The  most general $SU(3)\times SU(2) \times U(1)$ invariant scalar 
leptoquarks satisfying baryon and lepton number conservation have 
been written down Buchm\"uller {\sl et al.}\cite{buch}.  However, 
only those leptoquarks which couple to electrons can be produced in 
$e\gamma$ collisions and for real leptoquark production the chirality 
of the coupling is irrelevant.  For this case the number of 
leptoquarks reduces to four which can be distinquished by their 
electromagnetic charge; $Q_{em}= -1/3$, $-2/3$, $-4/3$, and $-5/3$.  
In our calculations we will follow the convention adopted in the 
literature\cite{leptoLN} where the leptoquark couplings are replaced 
by a generic Yukawa coupling $g$ which is scaled to electromagnetic 
strength $g^2/4\pi=\kappa \alpha_{em}$ and allow $\kappa$ to vary.

The process we are considering is shown if Fig. 1.  The parton level 
cross section is trivial, given by: 
\begin{equation}
\sigma(\hat{s})=\frac{\pi^2 \kappa \alpha_em}{M_s} 
                \delta(M_s - \sqrt{\hat{s}}).
\end{equation}
Convoluting the parton level cross section with the quark 
distribution in the photon one obtains the expression
\begin{eqnarray}
\sigma(s) & = & \int f_{q/\gamma}(z,M_s^2) \hat{\sigma}(\hat{s}) dz 
                \nonumber \\
& = & f_{q/\gamma}(M_s^2/s,M_s^2) 
      \frac{\mbox{$2\pi^2\kappa \alpha_{em}$}}{\mbox{$s$}}.
\end{eqnarray}
We note that the interaction Lagrangian used 
in Ref. \cite{HP} associates a factor $1/\sqrt{2}$ with the 
leptoquark-lepton-quark coupling.  Thus, one should compare our 
results with $\kappa$ to those in Ref. \cite{HP} with $2\kappa$.
We give results with $\kappa$ chosen to be 1.  

For $e^+e^-$ colliders
the cross section is obtained by 
convoluting the expression for the resolved photon contribution to 
$e \gamma$ production of leptoquarks, Eqn. (2), with the 
Weizs\"acker-Williams effective photon distribution
\begin{equation}
\sigma(e^+ e^- \rightarrow X S) = \frac{2 \pi^2 \alpha_{em}\kappa}{s} 
    \int_{M_s^2/s}^1 \frac{dx}{x} f_{\gamma/e}(x,\sqrt{s}/2) 
    f_{q/\gamma}(M_s^2/(x s), M_s^2)
\end{equation}
with the Weizs\"acker-Williams effective photon distribution given by 
\begin{equation}
f_{\gamma/e}(x,E) = \frac{\alpha_{em}}{2 \pi} \left\{ 
   \frac{[1 + (1 - x)^2]}{x} \ln \left[ \frac{E^2}{m_e^2} 
     \frac{(1 - 2 x + x^2)}{1 - x + x^2/4)} \right] 
   + x \ln \left( \frac{(2 - x)}{x} \right) 
   + \frac{2(x - 1)}{x} \right\}.
\end{equation}

There exist several different quark distribution functions in the 
literature \cite{nic,DO,DG,GRV,LAC}.  
The different distributions give almost identical results 
for the $Q_{LQ}=-1/3, \; -5/3$ leptoquarks  
and for the $Q_{LQ}=-2/3, \; -4/3$ leptoquarks 
give LQ cross sections that vary by most a factor 
of two, depending on the kinematic region.  We obtain our results using
the GRV distribution functions \cite{GRV} which we take to be 
representative of the quark distributions in the photon.

Before proceeding to our results we consider possible backgrounds 
\cite{DGP}.  The leptoquark signal consists of 
a jet and electron with balanced transverse momentum and 
possibly activity from the hadronic remnant of the photon.  
The only serious background is a hard scattering of a quark inside the 
photon by the incident lepton via t-channel photon exchange; $eq \to 
eq$.  We plot the invariant mass distribution for this background in 
our plots of the LQ cross sections and find that it is typically 
smaller than our signal by two orders of magnitude.
For the LQ invariant mass distribution we chose a 5~GeV invariant mass 
bin so that $d\sigma/dM =\sigma/5$~GeV.
Related to this process is the direct production of a 
quark pair via two photon fusion
\begin{equation}
e + \gamma \to e + q  + \bar{q}.
\end{equation}
However, this process is dominated by the collinear divergence which 
is actually well described by the resolved photon process $eq\to eq$ 
given above.  Once this contribution is subtracted away the remainder 
of the cross section is too small to be a concern \cite{DGP}.
Another possible background consists of $\tau$'s pair produced via 
various mechanisms with one $\tau$ decaying leptonically and the other 
decaying hadronically.  Because of the neutrinos in the final state it  
is expected that the electron and jet's $p_T$ do not in general 
balance which would distinguish these backgrounds from the signal.
However,  this background should be checked in a realistic 
detector Monte Carlo to be sure.
The remaining backgrounds originate from 
heavy quark pair production with one quark decaying 
semileptonically and only the lepton being observed with the 
remaining heavy quark not being identified as such. All such 
backgrounds are significantly smaller than our signal in the kinematic 
region we are concerned with.

In Fig. 2 we show the cross sections for $\sqrt{s}=91.17$~GeV.  
In our results we assume $BR(LQ \to e + q)=1$.  
We interpret the non-observation of any $e-jet$ events by any 
of the LEP experiments to signify that a LQ does not exist up to the 
leptoquark mass that would produce at least 3 $e-jet$ events
with the integrated luminosities of the combined LEP experiments.
Assuming $BR(LQ \to e \; jet)$ =1
this leads to  LQ search limits of 
$M_{LQ} > 90$~GeV for all first generation leptoquark types. 
If instead 
$BR(LQ \to e + q)=0.5 $ and $BR(LQ \to \nu + q)=0.5$ the second LQ 
decay mode would have an even more dramatic signature than the one we 
consider; a high $p_T$ monojet balanced against a large missing $p_T$.
Thus,  in this case the sum of the two possible decays would give 
limits similar to the one given above.

In Fig. 3 we show the cross sections for $\sqrt{s}=161$~GeV, the 
energy of the most recent LEP200 run.  For an integrated luminosity of 
10~pb$^{-1}$ per LEP experiment we interpret the non-observation of any 
$e-jet$ events for the LQ mass that would result in 3 events as scalar 
LQ search limits of $M_{LQ} > 130$~GeV for $Q_{LQ}=-2/3, \; -4/3$
and $M_{LQ} > 148$~GeV for $Q_{LQ}=-1/3, \; -5/3$ and 140~GeV and 154~GeV 
respectively for the vector leptoquarks.  These limits represent a 
significant improvement over the LEP limits given above and are
quite competive with similar published 
limits obtained by the Tevatron experiments \cite{D0,CDF}.  

Figures 4 and 5 give the cross sections for $\sqrt{s}=175$~GeV and
$\sqrt{s}=190$~GeV respectively.  At present we do not know for certain
what the total integrated luminosities will be for each of these 
energies.  We use the currently planned 25~pb$^{-1}$ at $\sqrt{s}=175$~GeV 
and 500~pb$^{-1}$ at $\sqrt{s}=190$~GeV for each of the LEP 
experiments.
Again, using the criteria that none of the experiments see any $e-jet$ 
events we obtain search limits up to $\sim$170~GeV for $\sqrt{s}=175$~GeV
and $\sim$188~GeV for $\sqrt{s}=190$~GeV. The search limits for the
different charge and spin assumptions are summarized in Table 1.

In this letter we have used the resolved photon contributions to 
single leptoquark production to estimate potential
limits on leptoquark masses.  
This production mechanism has the advantage that it can produce both 
scalar and vector leptoquarks with roughly the same cross section and 
of being insensitive to the chiral properties of the leptoquarks.  
It's only dependence on LQ properties is on the LQ charge because the 
the photon has a larger $u$-quark than $d$-quark content.  The 
potential limits 
we obtain from the non-observation of LQ's at $\sqrt{s}=161$~GeV are 
competative with published limits of the Tevatron collaborations.  In 
general, the limits obtainable at LEP200 will be slightly
lower than direct limits that can be 
obtained at HERA \cite{H1,ZEUS}.  Nevertheless, limits obtained in searches 
at LEP200 
should complement those obtained at HERA and the Tevatron for two 
reasons:  First, LQ's will have very dramatic 
and distinct signatures at LEP200 in contrast to the Tevatron and HERA 
where their presence are determined by statistical enhancements over 
the SM backgrounds.  Second, the limits obtained at LEP200 
tend to be less dependent on the properties of the LQ's than results 
obtained at $ep$ and hadron colliders.  Finally, we remind the reader 
that our results are of course only 
theorist's estimates which should be examined more closely and 
carefully than has been described here.  For example,  we have not 
included detector acceptances in our estimates, perhaps over 
simplistically assuming that the LQ decay products will decay 
isotropically in the detector resulting in fairly high detection 
efficiencies.  In addition, a detailed detector Monte Carlo of 
possible backgrounds should be performed.  Nevertheless we believe 
our estimates to be fairly 
robust and are not likely to be changed substantially by a more rigorous 
scrutinization.

\acknowledgments

This research was supported in part by the Natural Sciences and Engineering 
Research Council of Canada. The authors thank Bob Carnegie for helpful 
conversations and his encouragement and Dean Karlen for useful 
discussions.

\newpage
\begin{table}
\caption{Mass limits and discovery limits for leptoquarks at LEP and 
LEP200.  Our results are based on the non-observation of any 
leptoquark candidates when 3 are expected for the given LQ mass.  We 
take $\kappa=1$ and assume $BR(LQ \to e + q)=1$.  
Our results are given in GeV.}
\begin{tabular}{llcccc}
$\sqrt{s}$ & L & \multicolumn{2}{c}{Scalar LQ} &
			\multicolumn{2}{c}{Vector LQ} \\
(GeV) & (pb$^{-1}$) & $Q=-2/3 \; -4/3$ & $Q=-1/3 \; -5/3$
			& $Q=-2/3 \; -4/3$ & $Q=-1/3 \; -5/3$ \\
\tableline
91.17 & 720 & 90 & 90 & 90 & 91 \\
161 & 40 & 130 & 148 & 140 & 154 \\
175 & 100 & 154 & 167 & 162 & 170 \\
190 & 2000 & 187 & 188 & 188 & 189 \\
\end{tabular}
\end{table}

\newpage

\vskip 2.0cm
\centerline{\epsfig{file=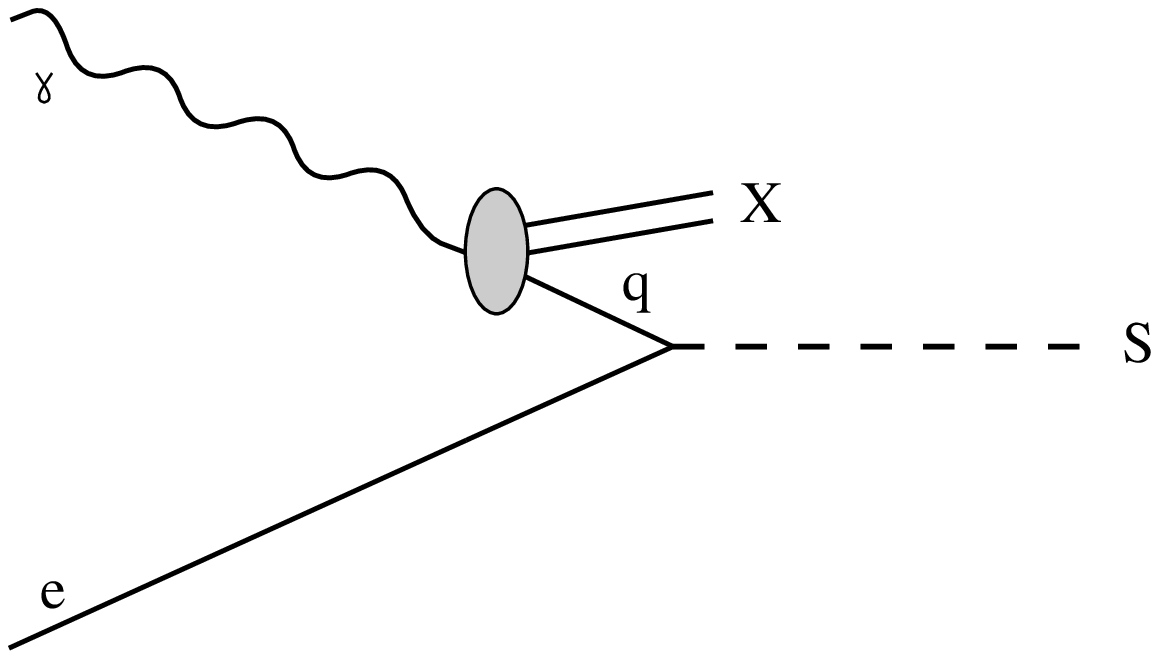,width=5.5cm,clip=}}
\noindent
{\bf Fig 1:} The resolved photon contribution for leptoquark production in 
$e\gamma$ collisions.

\vskip 3.0cm
\centerline{
\begin{turn}{90}
\epsfig{file=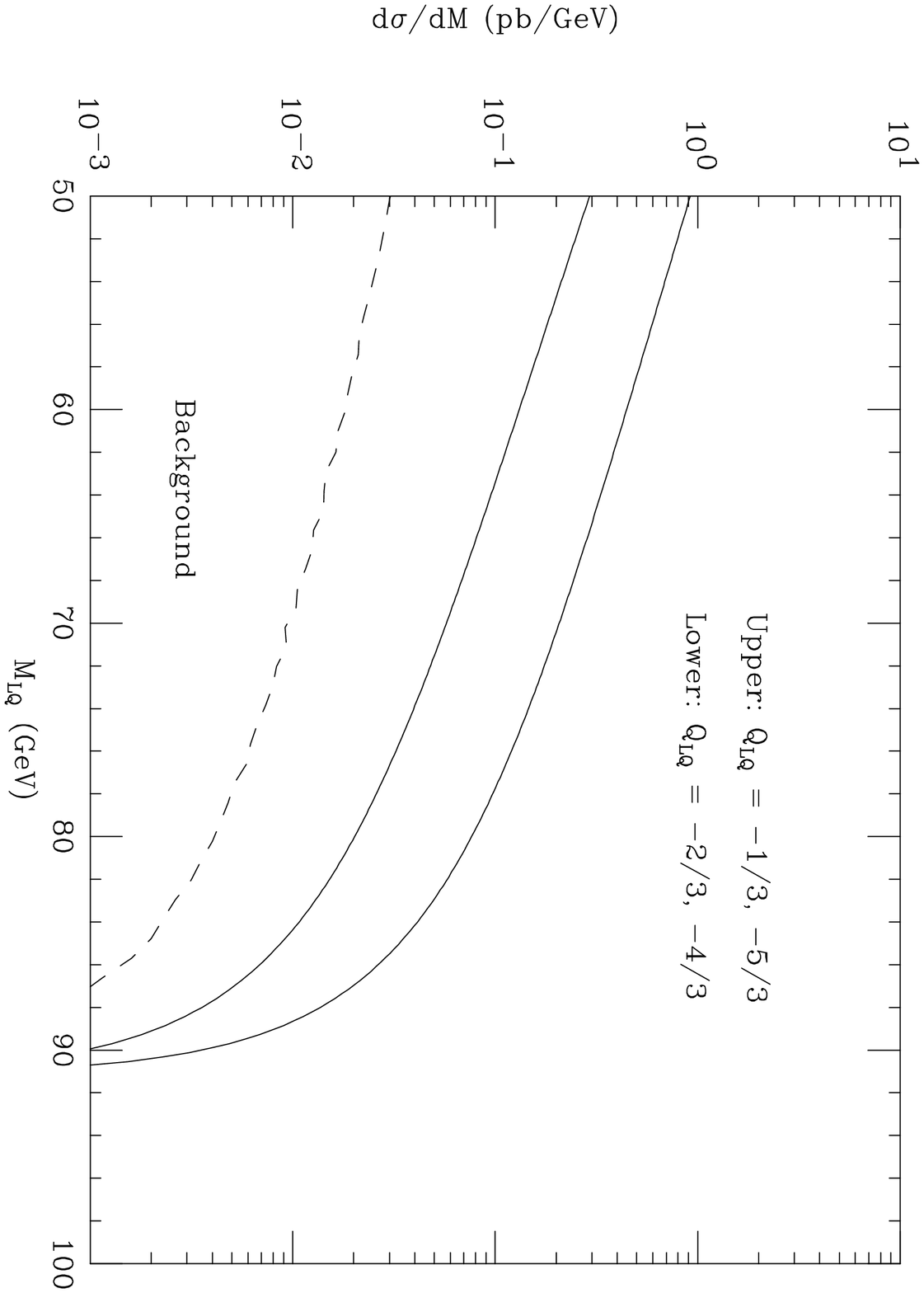,width=5.5cm,clip=}
\end{turn}
}
\noindent
{\bf Fig 2:}
The cross sections for scalar leptoquark production due to resolved 
photon contributions in $e^+ e^-$ collisions for $\sqrt{s}=91.17$~GeV. 
$\kappa$ chosen to be 1 and  resolved photon distribution 
functions of Gl\"uck, Reya and Vogt \cite{GRV}
are used. The dashed line is the $e[q]_\gamma \to e q$ background. 
For the LQ invariant mass distribution we use a 5~GeV invariant mass 
bin so that $d\sigma/dM =\sigma/5$~GeV. 

\newpage

\centerline{
\begin{turn}{90}
\epsfig{file=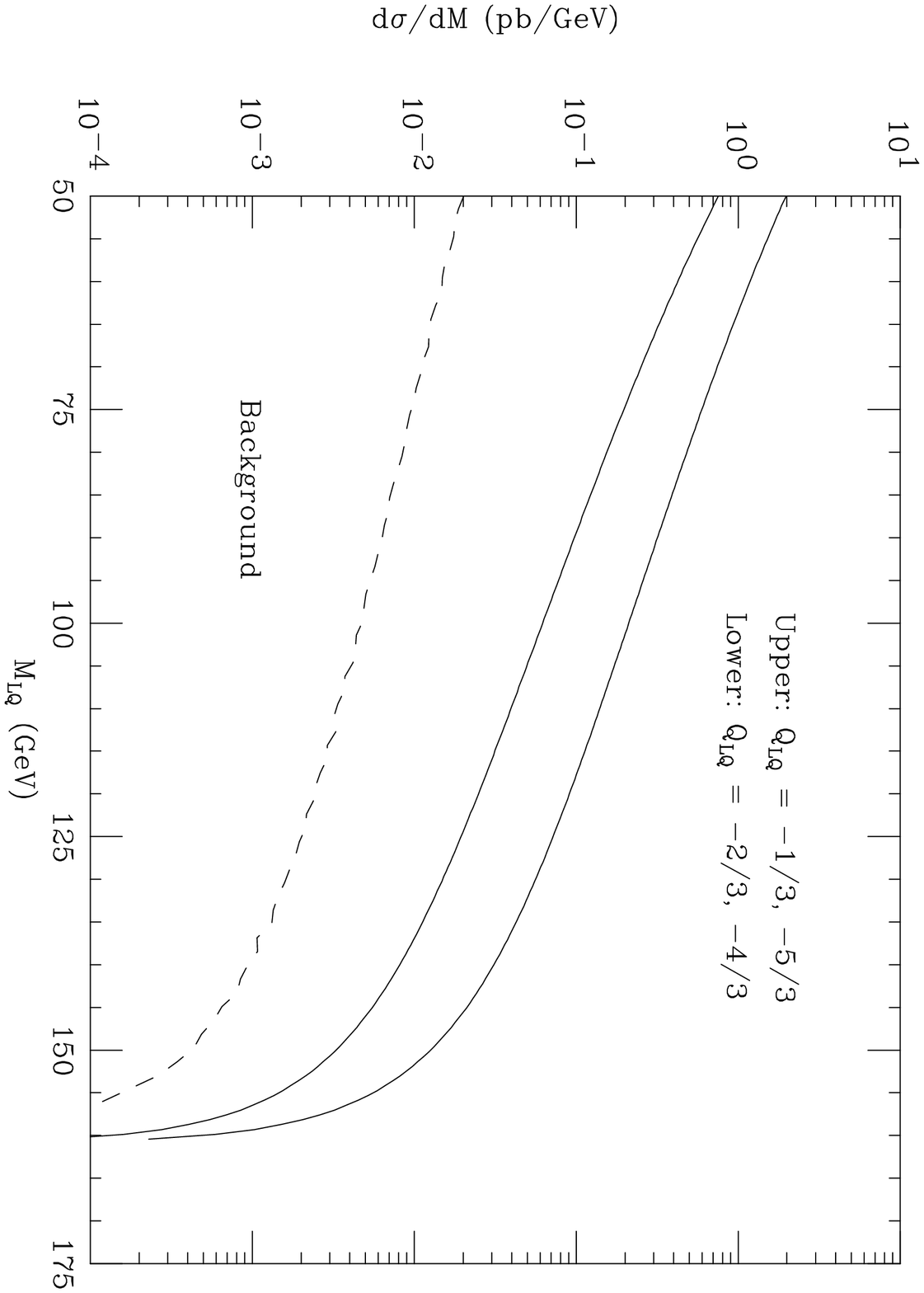,width=5.5cm,clip=}
\end{turn}
}
\noindent
{\bf Fig 3:} The cross sections for leptoquark production due to resolved 
photon contributions in $e^+ e^-$ collisions for $\sqrt{s}=161$~GeV.
See Fig. 2 for details.

\centerline{
\begin{turn}{90}
\epsfig{file=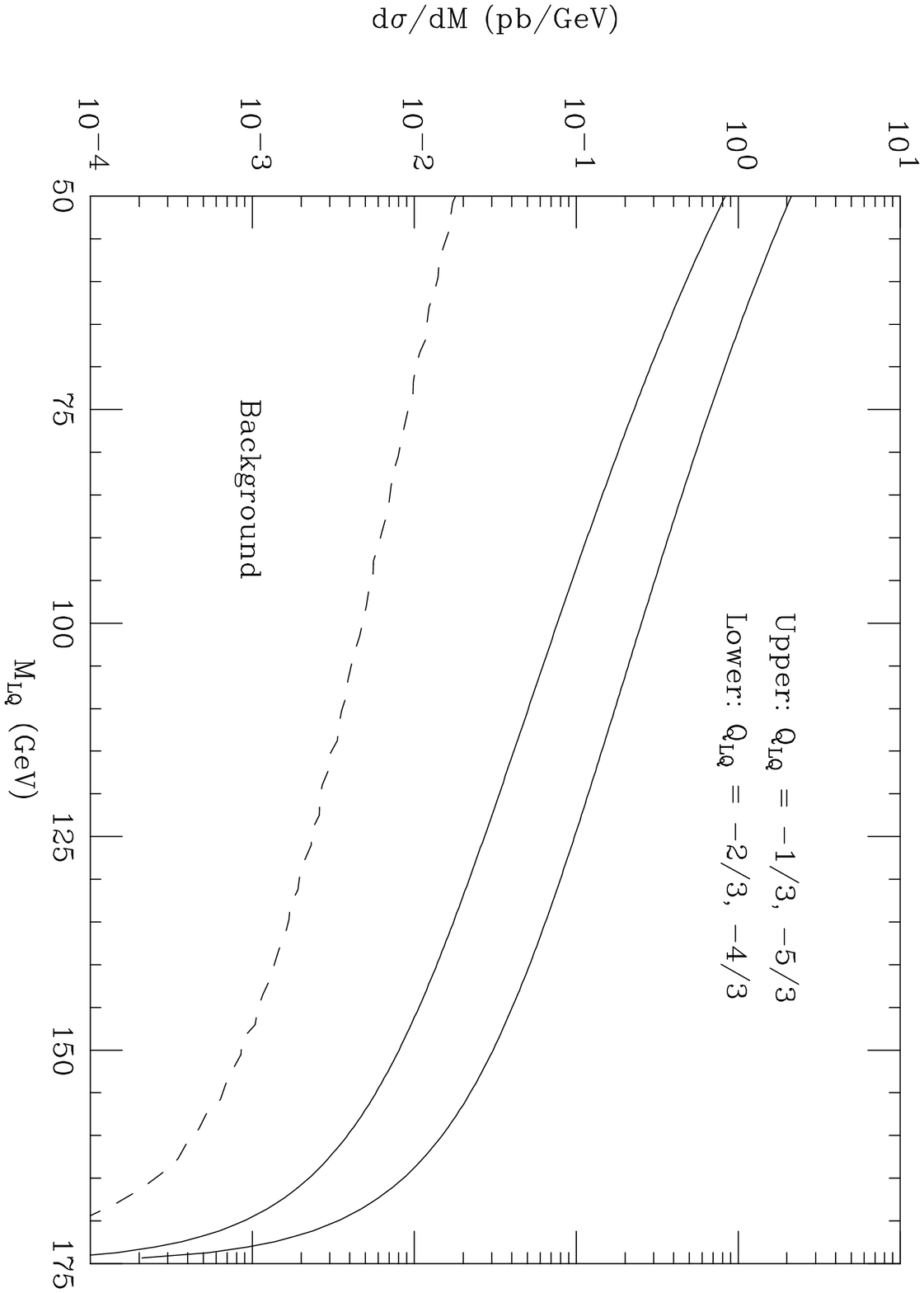,width=5.5cm,clip=}
\end{turn}
}
\noindent
{\bf Fig 4:} The cross sections for leptoquark production due to resolved 
photon contributions in $e^+ e^-$ collisions for $\sqrt{s}=175$~GeV.
See Fig. 2 for details.

\centerline{
\begin{turn}{90}
\epsfig{file=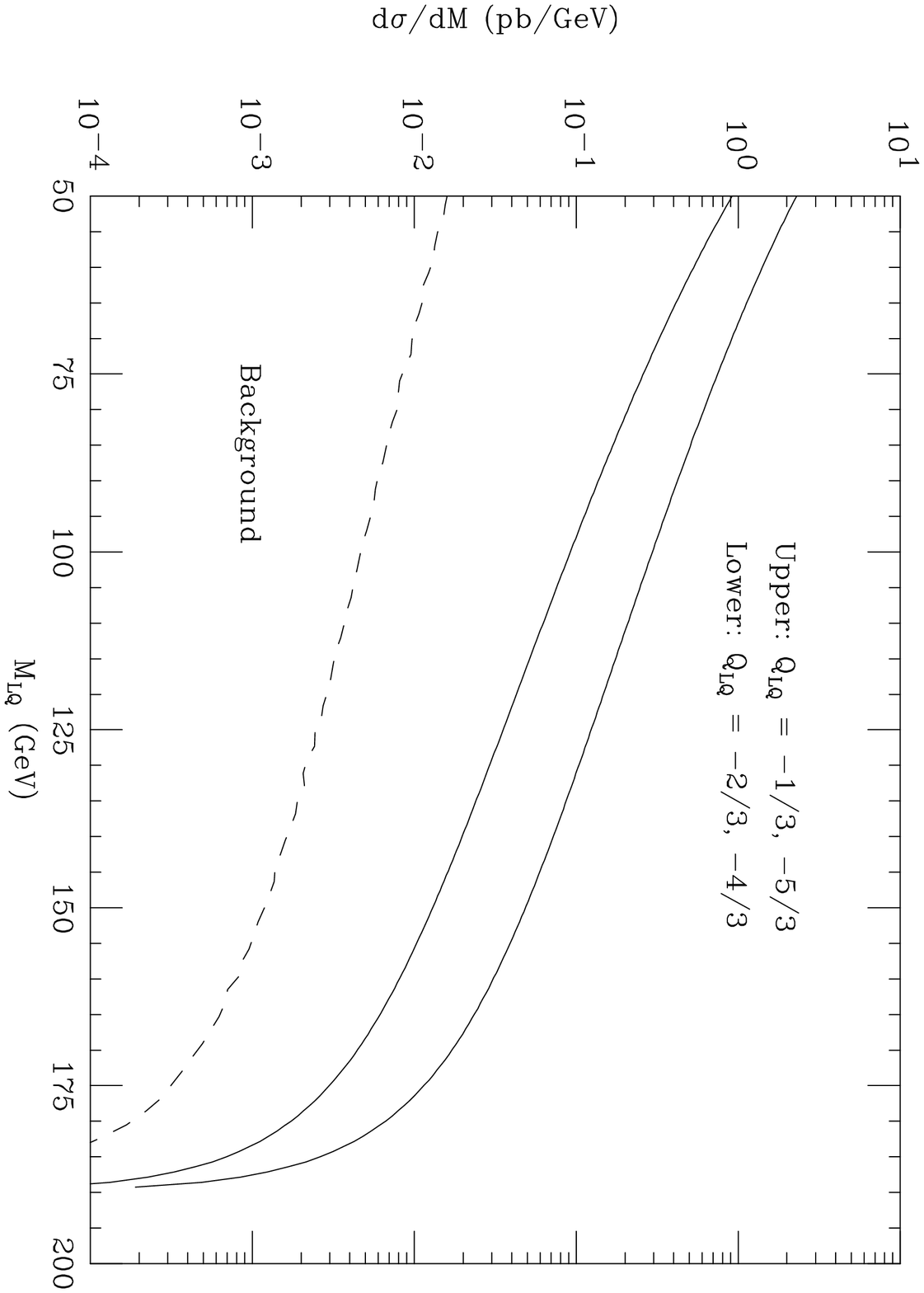,width=5.5cm,clip=}
\end{turn}
}
\noindent
{\bf Fig 5:}The cross sections for leptoquark production due to resolved 
photon contributions in $e^+ e^-$ collisions for $\sqrt{s}=190$~GeV.
See Fig. 2 for details.

\end{document}